\documentclass{aa}
\usepackage{natbib}
\bibpunct{(}{)}{;}{a}{}{,}
\usepackage{graphicx}
\usepackage[hang,bf,nooneline]{subfigure}
\usepackage[hang,bf,nooneline]{caption}
\usepackage{rotating}
\usepackage{pifont}
\usepackage{txfonts}

\begin{document}

\title{Determination of the Local Dark Matter Density in our Galaxy}

\author{M. Weber\inst{1} \and W. de Boer\inst{1}}

\offprints{Markus.Weber@ekp.uni-karlsruhe.de}

\institute{Institut f\"ur Experimentelle Kernphysik, Karlsruher Insitut 
           f\"ur Technologie (KIT), P.O. Box 6980, 76128 Karlsruhe, Germany}

\date{Received 30 September 2009 / Accepted 9 October 2009}

\authorrunning{M. Weber, W. de Boer}
\titlerunning{Constraints on Galactic Dark Matter}

\abstract{The rotation curve, the total mass and the gravitational
potential of the Galaxy are sensitive measurements of the dark matter
halo profile.}
{In this publication cuspy and cored DM halo profiles are analysed with
respect to recent astronomical constraints in order to constrain the shape of the
Galactic DM halo and the local DM density.
} {All Galactic density components (luminous matter and DM) are
parametrized. Then the total density distribution is
constrained by astronomical observations: 1) the total mass of
the Galaxy, 2) the total matter density at the position of the
Sun, 3) the surface density of the visible matter, 4) the
surface density of the total matter in the vicinity of the Sun,
5) the rotation speed  of the Sun and 6) the shape of the
velocity distribution within and above the Galactic disc. The
mass model of the Galaxy  is mainly constrained by the local
matter density (Oort limit), the rotation speed of the Sun and
the total mass of the Galaxy from tracer stars in the halo. }
{It is shown from a statistical $\chi^2$ fit to all data that
the local DM density is strongly positively (negatively)
correlated with the scale length of the DM halo (baryonic
disc). Since these scale lengths are poorly constrained the
local DM density can vary from 0.2 to 0.4 GeV cm$^{-3}$
(0.005 - 0.01 M$_\odot$ pc$^{-3}$) for a
spherical DM halo profile and allowing total Galaxy masses up
to 2 $\cdot$ 10$^{\mathrm{12}}$ M$_\odot$. For oblate DM halos and dark
matter discs, as predicted in recent N-body simulations, the
local DM density can be increased significantly. } {}

\keywords{---Milky Way: halo, structure, dark matter, rotation
curve, dwarf galaxies, dark matter mass, baryonic mass ---Cosmology:
dark matter, dwarf galaxies, structure formation}

\maketitle

\section{Introduction}
\label{sec:Introduction}

The best evidence for Dark Matter (DM) in galaxies is usually provided by rotation
curves, which do not fall off fast enough at large distances from the centre.
This can  be understood, either by assuming Newton's law of gravitation is not
 valid at large distances, the so-called MOND (Modified Newtonian Dynamics) theory
\citep{Bekenstein:1984tv,Bienayme:2009wb}, or the visible mass distribution
is augmented by invisible mass, i.e. DM.
For a flat rotation curve the DM density has to fall off like 1/r$^2$ at
large distances.

Given the overwhelming evidence for DM on all scales from the flatness of the universe
combined with gravitational lensing and structure formation we assume DM exists and
try to constrain the DM density from dynamical constraints, for which better data
became available in recent years:
\begin{enumerate}
  \item the total mass of the Galaxy has to be about 10$^{12}$ solar masses \citep{Wilkinson:1999hf,
	Battaglia:2005rj,Xue:2008se};
  \item the total mass inside the solar orbit is constrained by the well-known
        rotation speed of the solar system;
  \item the total matter density at the position of the Sun from the gravitational
        potential determined from the movements of local stars as measured with the
        Hipparcos satellite \citep{Holmberg:2004fj};
  \item the surface density of the visible matter at the position of the Sun \citep[and references therein]{Naab:2005km};
  \item the surface density of the total matter at the position of the Sun \citep{Kuijken:1991mw,
        Holmberg:2004fj,Bienayme:2005py};
  \item the shape of the rotation curve within Galactic disc \citep[and references therein]{Sofue:2008wt};
  \item the  velocity distribution above the Galactic disc
      ($z>4$ kpc) \citep{Xue:2008se}.
\end{enumerate}
Towards the Galactic centre (GC) N-body simulations of
structure formation predict a DM density, which diverges at the
centre, thus developing a so-called cusp, with the slope
varying more like 1/r than 1/r$^2$
\citep{Navarro:1996gj,Moore:1999nt,Ricotti:2002qu}. On the
contrary,  rotation curves of nearby dwarf galaxies do not show
such a cusp, but rather a constant density near the center, a
so-called core \citep{Oh:2008ww,Gentile:2007sb,Salucci:2007tm}. 
N-body simulations including
the complicated baryonic disc formation indicated that the
cusps may have been destroyed by the fluctuation in the
gravitational potential of the baryons, which is much stronger
than the potential caused by the DM in the centre
\citep{Mashchenko:2006dm}. In addition, the DM may form a disc
like structure by the infall of DM dwarf galaxies, as shown by
recent high resolution N-body simulations
\citep{Purcell:2009yp}. Evidence for substructure of DM inside
the disc is supported by the structure in the gas flaring
\citep{Kalberla:2007sr}, by the peculiar change of slope for
the rotation curve inside the disc, which is not observed for
stars outside the disc (Sect. \ref{subsec:RC}) and by the
structure in the diffuse gamma radiation \citep{deBoer:2005tm}.

Such substructure will not be investigated in this paper, but
smooth DM halos with different (cored and cuspy) profiles will
be compared with all available data. These will result in lower
limits on the local DM density, since  dark matter discs or
other local substructure will only enhance the local density.

A reliable determination of the local DM density is of great interest for
direct DM search experiments, where elastic collisions between WIMPs and
the target material of the detector are searched for. This signal is proportional
to the local density. A review on direct searches can be found in the paper
by \citet{Spooner:2007zh}.

The structure of the paper is as follows:  in section \ref{sec:para} the
parametrization of the luminous matter and five different DM halo profiles
are given. In Sect. \ref{sec:data} the experimental
 data used to determine the mass model are discussed. Then the
numerical determination of the mass model parameters using a $\chi^2$ fit
of all dynamical constraints is discussed in Sect. \ref{sec:fit}.
At the end a summary of the results is given.


\section{Parametrization of the Density Distributions}
\label{sec:para}
In order to constrain the mass model of the Galaxy by data it is convenient
to have a parametrization for both, the visible an dark matter density.
These parametrizations are introduced here.
\subsection{Parametrization of the Luminous Matter Density}
\label{subsec:para_lumi}
The density distribution of
the luminous matter of a spiral galaxy is split into two parts,
the Galactic disc and the Galactic bulge. The parametrization of
the density distribution of the bulge is adapted from
the publication by \citet{Cardone:2005qq}
\begin{eqnarray}
  \rho_{b}(r,z) & = & \rho_b \cdot \left( \frac{m}{r_{0,b}}
  \right)^{-\gamma_b} \cdot \left( 1 + \frac{\tilde{r}}{r_{0,b}} \right)^{\gamma_b -
  \beta_b} \exp{\left( -\frac{\tilde{r}^2}{r_t^2}\right)},\\
  \nonumber \tilde{r}^2 & = & \sqrt{x^2+y^2 + (z/q_b)^2}.
\label{eq:bulge}
\end{eqnarray}
For a good description of the RC near the GC the
parameters of the bulge profile are found to be $\gamma_\mathrm{b} = \beta_\mathrm{b} =$ 1.6,
q$_\mathrm{b} =$ 0.61, r$_\mathrm{t} =$ 0.6 kpc, r$_{\mathrm{0,b}}$ $=$ 0.7 kpc
and $\rho_\mathrm{b} =$ 360 GeV cm$^{-3}$,
which corresponds to approximately $9.5$ M$_\odot$ pc$^{-3}$. This high
density might be influenced by the presence of black holes.
At least one black hole with a mass of about $4.0 \cdot 10^6$ M$_\odot$
has been observed in the GC by \citet{Reid:2008fp}.\\
The stellar contribution of the Galactic disc is split into two
discs - a {\it thin} and a {\it thick disc} - which are usually
parametrized by an exponentially decreasing density distribution.

The parametrization of the Galactic disc is taken from the publication
by \citet{Sparke}
\begin{equation}
  \rho_{d}(r,z) = \rho_d \cdot exp(-r/r_d) \cdot exp(-z/z_d).
\label{eq:disc}
\end{equation}
The parameter $\rho_\mathrm{d}$ describes the density of the Galactic disc at the
GC while r$_\mathrm{d}$ and z$_\mathrm{d}$ describe the scale parameter in radial and vertical
direction.

There is some freedom in the
choice of the parameters for the Galactic disc. Its density in the
GC is, as in case of the bulge, unknown, so it has to be a free
parameter. For the scale radius we adopt the value from \citet{Hammer:2007ki}
\begin{equation}
 r_d=2.3\pm0.6 \ \mathrm{kpc}.
 \label{rd}
 \end{equation}
The scale height z$_\mathrm{d}$ varies for the different components: young stars
are born near z = 0, so they have a much smaller scale height, while
the older stars have been kicked around and reach scale heights up
to 700 pc.
The disc has two distinct populations with two different scale heights.
The thin disc consists
of young, bright and metal-rich stars which provide about 98\% of
the total stellar population \citep{Gilmore:1983bv}. Its scale height
was determined from star counts of stars with an absolute magnitude
of M$_\mathrm{V}$ $>$ 6 to be z$_\mathrm{d} \approx$ 270 pc \citep{Kroupa:1993ga}.
From measurements of the spatial distribution of K dwarf stars the
vertical scale height turns out to be smaller than 250 pc
\citep{Robin:1995fr}. In the publication by \citet{Ojha:1995ie}
the scale height of the thin disc was determined for stars with
M$_\mathrm{V} \gtrsim$ 3.5 to be 260 $\pm$ 50 pc. Furthermore the interstellar
gas and dust contribution is also located in the thin disc. The
thick disc consists of old, metal-poor stars and could be the result
of either an earlier thin disc or the merging of a satellite galaxy
with the MW \citep{Robin:1995fr}. Its scale height lies between
700 and 1500 pc and its local density is about 5\% of the local
stellar density of the thin disc. For that reason in this
analysis the luminous matter is given only by the thin disc
while small density contribution of the thick disc is neglected.
For gas, which makes up 10\% of the mass of the disc, the scale
height varies with Galactocentric radius because of the decreasing
gravitational potential at larger radii. Fortunately, the mass model
of the Galaxy is not very sensitive to the exact value of the scale
height, so it is fixed at 320 pc, which is close to the value adopted
by \citet{Freudenreich:1997bx}.
The mass of the bulge was taken into account, but the detailed 
structure of the mass distribution (hole in the centre of the disc \citet{Freudenreich:1997bx}, the
bar structure and the black hole) are not taken into account in the
parametrization, since the para\-meters of interest, i.e. the DM halo
parameters, are insensitive to the detailed mass distribution in the
centre of the Galaxy, since the gravitational potential in the centre is completely
dominated by the baryonic matter.

The  parametrization of the visible mass discussed above
leads to a mass of the Galactic bulge of about 1.1 $\cdot$ 10$^{\mathrm{10}}$ M$_\odot$.
The mass of the Galactic disc varies for different fits because of the
variation of the parameters $\rho_\mathrm{d}$ and r$_\mathrm{d}$. It is in the range of
5 $\cdot$ 10$^{\mathrm{10}}$ to 7 $\cdot$ 10$^{\mathrm{10}}$ solar masses.
In addition to the luminous matter the density profile of the DM halo
has to be parametrized. This is discussed in the following section.

\subsection{Parameterization of the Dark Matter density}
\label{subsec:para_dm}

The first analytical analysis of
structure formation in the Universe by \citet{Gunn:1977mw}
predicted that DM in Galactic haloes are distributed according to a
simple power law distribution $\rho(\mathrm{r}) \propto \mathrm{r}^{-\gamma}$. However,
later studies based on numerical N-body simulations \citep{Navarro:1996gj,Moore:1999nt}
found that the slope of the density distribution in the DM halo is
different for different distances from the GC. Today
it is commonly believed that the profile of a DM halo can
be well fitted by the universal function
\begin{eqnarray}
  \rho_\chi(r) & = & \rho_{\odot,DM} \cdot \left ( \frac{\tilde r}{r_\odot} \right )^{-\gamma}
  \cdot \left \lbrack \frac{1 + \left ( \frac{\tilde r}{a} \right )^\alpha}{1 + \left
  ( \frac{r_\odot}{a} \right )^\alpha} \right \rbrack^{\frac{\gamma -
  \beta}{\alpha}},\\
  \nonumber \tilde r & = & \sqrt{x^2 + \frac{y^2}{\epsilon_{xy}^2} + \frac{z^2}{\epsilon_z^2}}.
\end{eqnarray}
Here, a is the scale radius of the density profile, which determines
at what distance from the centre the slope of the profile changes,
$\epsilon_{\mathrm{xy}}$ and $\epsilon_{\mathrm{z}}$ are the eccentricities of the
DM halo within and perpendicular to the Galactic plane and r$_\odot$
is the Galactocentric distance of the solar system. In the rest of
this paper we will assume axisymmetrical haloes, i.e.
$\epsilon_{\mathrm{xy}}=$ 1, but $\epsilon_\mathrm{z}$ is allowed to have values below
or above 1, thus allowing oblate and prolate haloes.
The parameter $\rho_{\odot,\mathrm{DM}}$ is the DM density of the halo at
the position of the Sun.
The parameters $\alpha$, $\beta$ and $\gamma$ characterize the radial
dependence of the density distribution. For r $\approx$ a the slope
of the halo profile is about r$^{-\alpha}$, for r $>>$ a  $\approx$ r$^{-\beta}$ and
for r $<<$ a  $\approx$ r$^{-\gamma}$.
Different sets for these parameters were suggested by different
analyses. Generally the results from numerical simulations show
that the DM density in the GC is divergent. Such profiles are called
"cuspy" profiles due to the cusp in the centre. In this analysis
three cuspy profiles are considered. In the publication by
\citet{Navarro:1996gj} it was found that their simulation results can be
approximated by a profile with a slope of $\gamma =$ 1 in the
GC, a stronger decreasing slope of $\beta =$ 3 for large radii and $\alpha =$ 3 for
the intermediate distances. This profile will be referred to
as NFW.
From later simulations by
\citet{Moore:1999nt} a profile with a somewhat steeper slope
of $\gamma =$ 1.5 in the GC (hereafter Moore) was favoured. A third cuspy halo
profile was introduced in \citep{Binney:2001wu}. They found that
the microlensing data towards the centre of the MW produced by
the MACHO collaboration is compatible with a density profile with
$\gamma \gtrsim$ 0.3 (hereafter BE).
Also recent N-body simulations showed a less cuspy profile for the clumpy
component of the DM, which is usually called the Einasto profile \citep{Ludlow:2008qf}.
The amount of matter in clumps is uncertain, but usually considered small compared
with the diffuse component \citep{Springel:2008by,Springel:2008zz}.
By considering cusps with slopes between 0.3 and 1.5
one probably covers the range of possible slopes for cuspy profiles.

\begin{figure}[tbp]
\begin{center}
\includegraphics[width=0.5\textwidth]{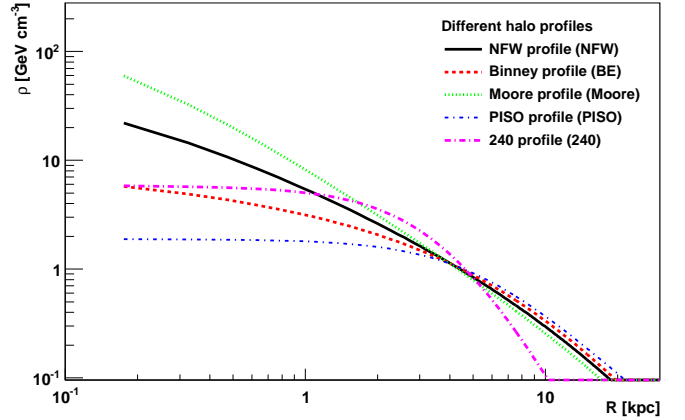}
\caption{Radial DM density distribution for the parameters
         given in Table \ref{tab:profiles}. The normalization $\rho_\mathrm{0}$ of
         the halo profiles is fixed for each profile by the requirement
         of the rotation speed of the solar system. }
\label{fig:halo_para}
\end{center}
\end{figure}

In contrast to cuspy profiles
the density distributions preferred by observations of rotation curves
of low surface brightness galaxies and dwarf spiral galaxies have a
nearly constant DM density in the GC  ($\gamma \approx$ 0) 
\citep{Oh:2008ww,Gentile:2007sb,Salucci:2007tm}.
Such profiles are called "cored" profiles
due to the constant density in the kpc scale in the central region.
Two different cored halo
profiles are considered in this analysis. The first profile is called
pseudo-isothermal profile (hereafter PISO) since it is an isothermal profile
($\propto$ r$^{-2}$) which is flattened in the centre.
The second cored halo
profile (hereafter 240) is similar to the PISO profile but decreases
faster for large radii. The parameter settings
of the several density profiles are shown in Table \ref{tab:profiles}.
The local DM density $\rho_{\odot,\mathrm{DM}}$ is a priori an unknown parameter
and therefore a degree of freedom in the density model.
\begin{table}[tpb!]
  \begin{center}
    \caption{Parameter settings for the different DM halo profiles
             considered in this analysis. The parameters of the NFW profile
             are adapted from the publication by \citet{Navarro:1996gj} but in this
             analysis a larger scale radius a is used.}
    \begin{tabular}{c|c|c|c|c|c}\hline
      Profile & $\alpha$ & $\beta$ & $\gamma$ & a [kpc] & Reference \\\hline
      NFW & 1.0 & 3.0 & 1.0 & 20.0 & \citet{Navarro:1996gj}\\
      BE & 1.0 & 3.0 & 0.3 & 10.2 & \citet{Binney:2001wu}\\
      Moore & 1.5 & 3.0 & 1.5 & 30.0 & \citet{Moore:1999nt}\\
      PISO & 2.0 & 2.0 & 0.0 & 5.0 & \citet{deBoer:2005tm}\\
      240 & 2.0 & 4.0 & 0.0 & 4.0 & \\\hline
    \end{tabular}
    \label{tab:profiles}
  \end{center}
\end{table}
In the publications by \citet{Amsler:2008zzb,Gates:1995dw} its value
is quoted to be in the range 0.2 - 0.7 GeV cm$^{-3}$ (0.005 - 0.018 M$_\odot$ 
pc$^{-3}$). In this ana\-lysis $\rho_{\odot,\mathrm{DM}}$ is left free. In
Fig. \ref{fig:halo_para} the different profiles are shown for equal
masses within the solar orbit.

\section{Dynamical Constraints on the Mass Model of the Milky Way}
\label{sec:data}
As mentioned in the introduction the mass model of the Galaxy is
largely constrained by the total mass of the Galaxy, the rotation
velocity of the Sun at its orbital radius r$_\odot$ and the local matter
density $\rho_{\odot,\mathrm{tot}} = \rho_{\mathrm{tot}}$(r$_\odot$), known as
the Oort limit. An additional constraint comes from the local
surface density. The experimental input for these constraints
is discussed first.

\subsection{Total mass}
\label{subsec:massdistribution}
The total mass of the MW is an important quantity in order
to constrain the DM density distribution. In general, it is
measured indirectly either via the kinematics of
distant halo tracer stars or satellite galaxies or the vertical scale
height of the gas distribution of the Galactic disc, which can
be measured at large distances from the GC.

The definition of the total mass of the Galaxy is difficult since a
slowly decreasing density has
an infinite extension. The total mass
of a galaxy is conventionally defined as the mass within the so-called
{\it virial radius}. At this radius the total mass of the accumulated
density of the MW is equal to the mass of a homogeneous sphere with the
constant density of 200 times the critical density of the Universe.

In the paper by \citet{Wilkinson:1999hf} the mass of the MW was
estimated from measurements of the radial velocities of 27
globular clusters and satellite galaxies for Galactocentric
distances R $>$ 20 kpc, using a Bayesian likelihood method and
a spherical halo mass model with a truncated radius. They found
a mass of the Galaxy within 50 kpc of M(50 kpc) $=
5.4^{+0.2}_{-3.6} \cdot 10^{11}$ M$_\odot$ and a total mass of
M$_{\mathrm{tot}} = 1.9^{+3.6}_{-1.7} \cdot 10^{12}$ M$_\odot$. A similar
ana\-lysis was done with more tracer stars by
\citet{Sakamoto:2002zr}; they find M$_{\mathrm{tot}} = 2.5^{+0.5}_{-1.0}
\cdot 10^{12}$ M$_\odot$. These measurements used a  simple
parametrization of the potential. Analyses using an NFW profile
for the DM distribution usually find a lower total mass, given
it steeper fall-off of the density profile at large distances.

Using a large sample of 2400 BHB stars from the SDSS in the
halo (z $>$ 4 kpc, R$<60$ kpc) and comparing the results with
N-body simulations using an NFW profile \citet{Xue:2008se} find
\begin{equation}
\mathrm{M}_{\mathrm{R}<60~{\mathrm{kpc}}} =4.0 \pm 0.7 \cdot 10^{11} \mathrm{M}_\odot ,
\label{mtot}
\end{equation}
which corresponds to M$_{\mathrm{tot}} = 1.0^{+0.3}_{-0.2} \cdot
10^{12}$ M$_\odot$. To get this total mass adiabatic contraction
\citep{Blumenthal:1985qy}  was allowed in which case the
concentration parameter was taken as a free parameter in the
fit. This parameter came out to be low (c $=$ 6.6$^{+1.8}_{-1.6}$;
if adiabatic contraction is ignored, the concentration
parameter becomes around 12 and the mass decreases to M$_{\mathrm{tot}}$ 
$=$ 0.9 $\cdot$ 10$^{\mathrm{12}}$ M$_\odot$, which is well within the errors.
Similar  mass values were found from the velocity dispersion by
\citet{Battaglia1:2005rj}, although with larger errors. These
measurements are consistent with the favoured $\Lambda$CDM
model in \citet{Klypin:2001xu} where the virial mass is $1
\cdot 10^{12}$ M$_\odot$. The value from Eq. \ref{mtot} will be
used in the analysis.

\begin{figure}[tbp]
\begin{center}
\includegraphics[width=0.5\textwidth]{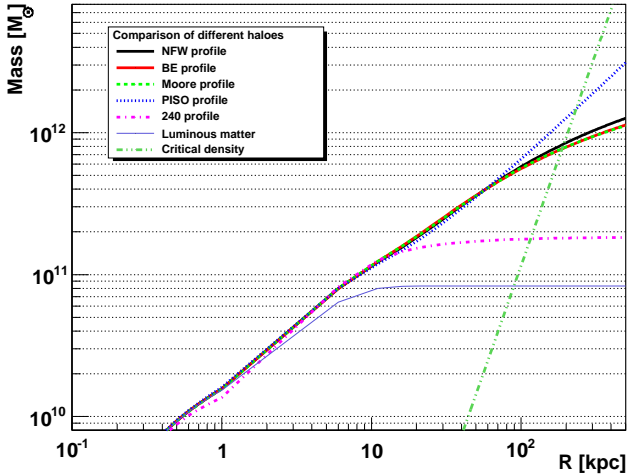}
\caption{The mass inside a radius as function of that radius is shown for the different
         halo profiles defined in Table \ref{tab:profiles}.
	 The thin solid line represents the visible mass which is different for
	 different halo profiles because of the variation of the parameters $\rho_\mathrm{d}$ and
	 r$_\mathrm{d}$. Here the luminous matter for the NFW profile is shown.
         The steep line starting at 40 kpc represents the mass of a sphere with a density
         of 200 times the critical density of the Universe.
         The crossings of the mass distributions
         with this line represent the virial radius and the total
         Galactic mass of the corresponding density distribution.}
\label{fig:mass_simple}
\end{center}
\end{figure}

Fig. \ref{fig:mass_simple} shows the radial dependence of the
total Galactic mass for the different spherical halo profiles. At small
radii r $<$ 5 kpc the density distribution is dominated by the luminous
matter, shown by the thin solid line, which is independent of the halo profile.
 The mass of a homogeneous sphere
with 200 times the critical density of the Universe is also shown. The
crossing of a mass distribution with this line defines the total Galactic
mass and the virial radius for this density distribution which is 210 kpc
for a Galactic mass of 10$^{12}$ M$_\odot$.\\
The mass distributions of the NFW and the BE profile are quite
similar since these profiles  differ only in the region around
the GC where the influence of DM is small. The 240 profile yields the smallest
mass while the mass distribution of the PISO profile shows a linear
increase with radius. The reason is the quadratic decrease
($\propto$ r$^{-2}$) of the PISO profile. Consequently, the integral
of such a profile leads to a linear increase of the Galactic mass.
By going from spherical to elliptical profiles the mass
can be changed significantly, as will be discussed later.

\begin{figure*}[tbp]
  \begin{center}
    \subfigure[Rotation curve ( z $=$ 0 kpc ), non-averaged]{\includegraphics[width=0.49\textwidth,angle=0]{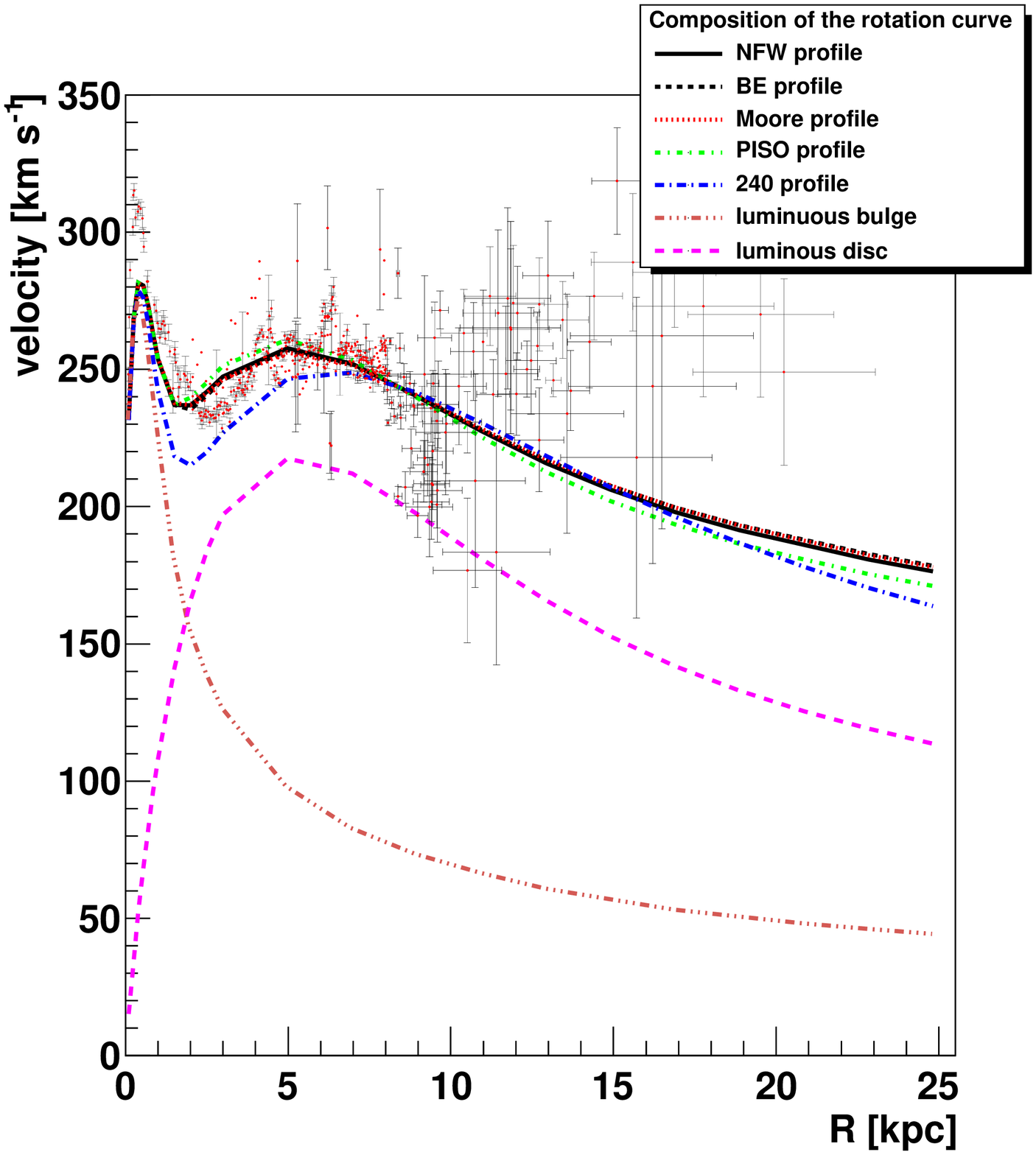}}
    \subfigure[Rotation curve ( z $=$ 0 kpc ), averaged]{\includegraphics[width=0.49\textwidth,angle=0]{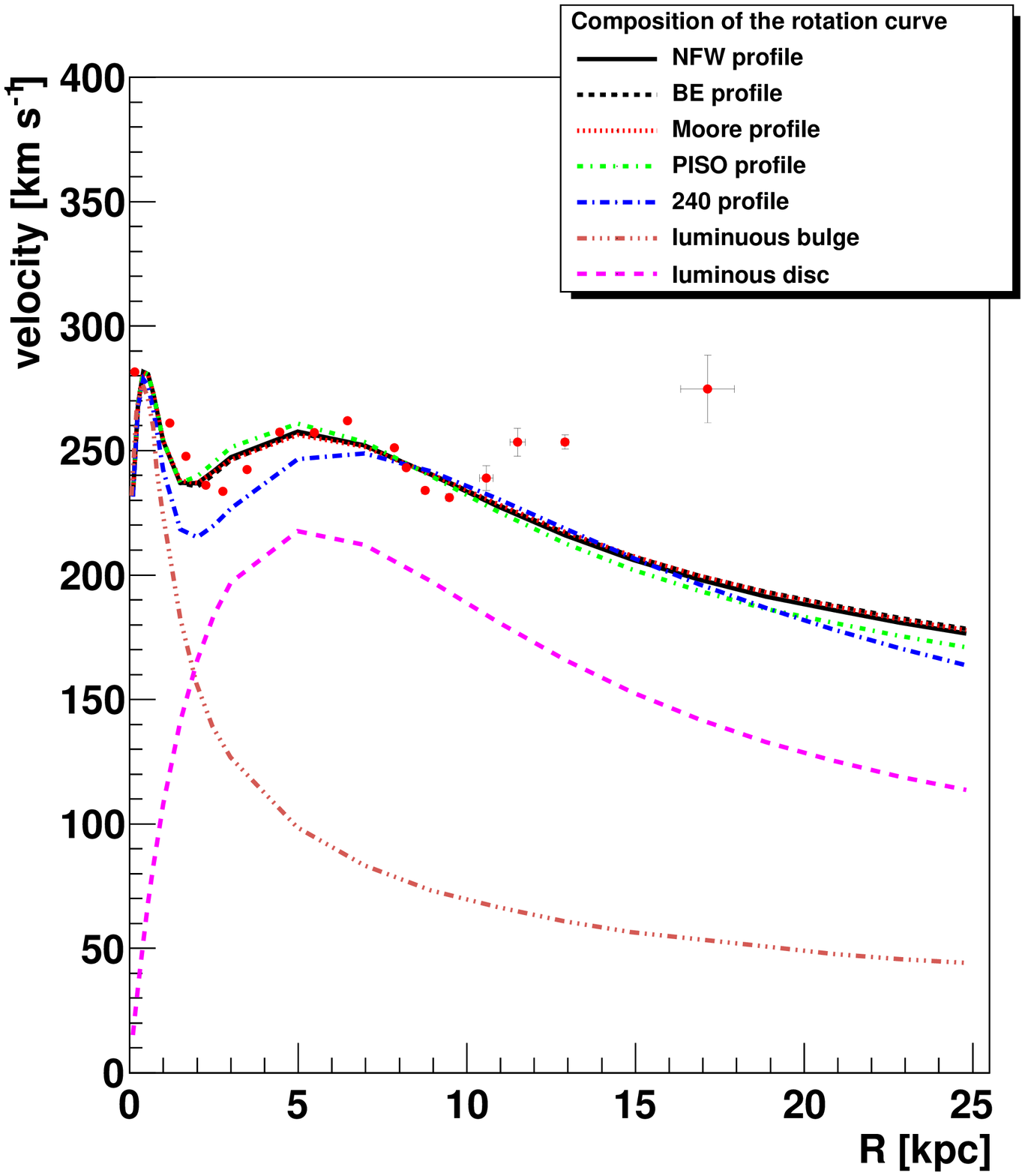}}
    \caption{The rotation curves - calculated for different halo profiles - in comparison with experimental data,
             which have been adapted from the publication by
             \citet{Sofue:2008wt}. On the left side all data points are shown, while on the right side a weighted average of
             the experimental data is shown  in 17 radial bins.}
    \label{fig:rc_simple}
  \end{center}
\end{figure*}
\begin{figure*}[tbp]
  \begin{center}
    \subfigure[Velocity curve (z $>$ 4 kpc), NFW profile]{\includegraphics[width=0.49\textwidth,angle=0]{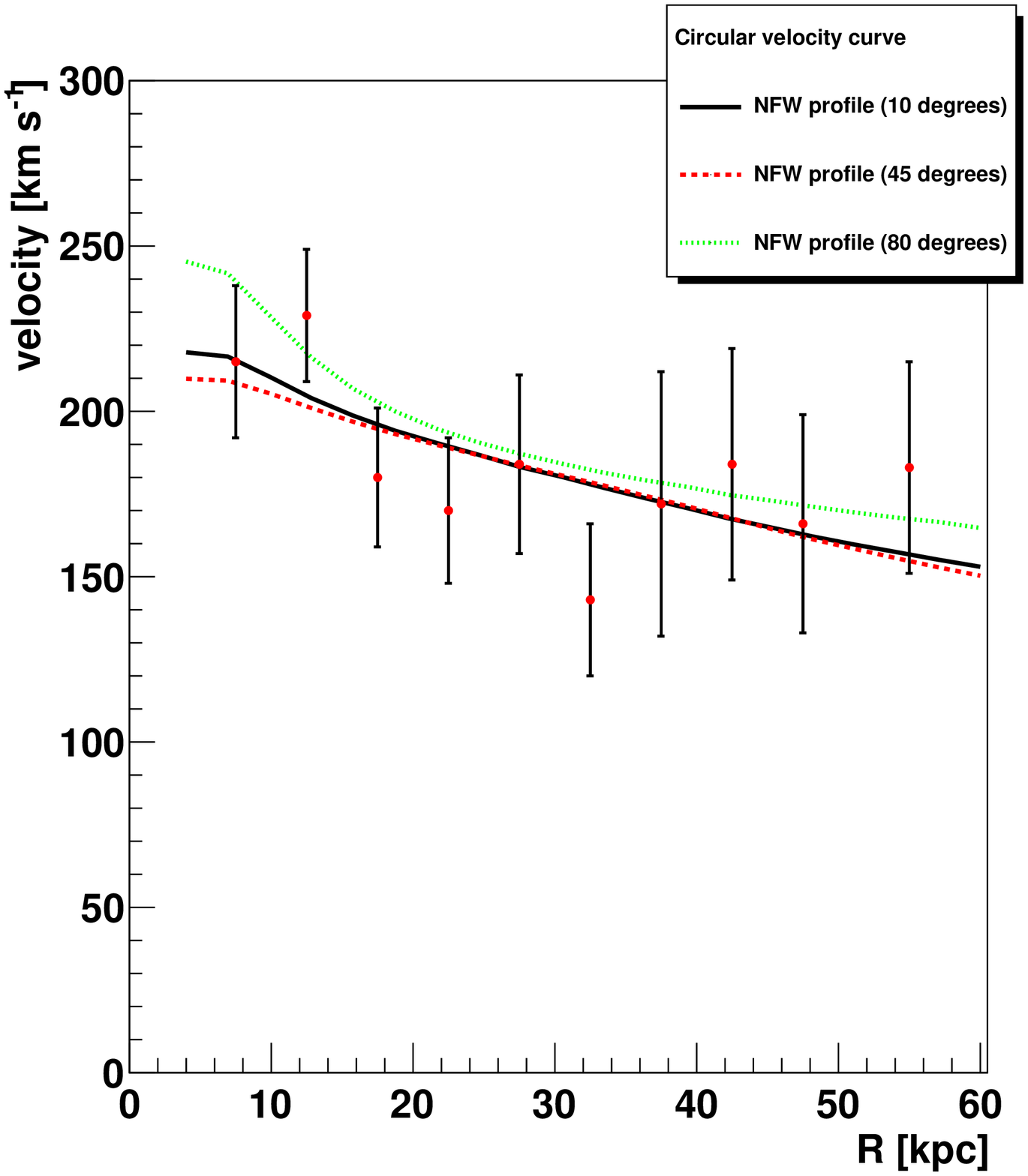}}
    \subfigure[Velocity curve (z $>$ 4 kpc) for different DM profiles]{\includegraphics[width=0.49\textwidth,angle=0]{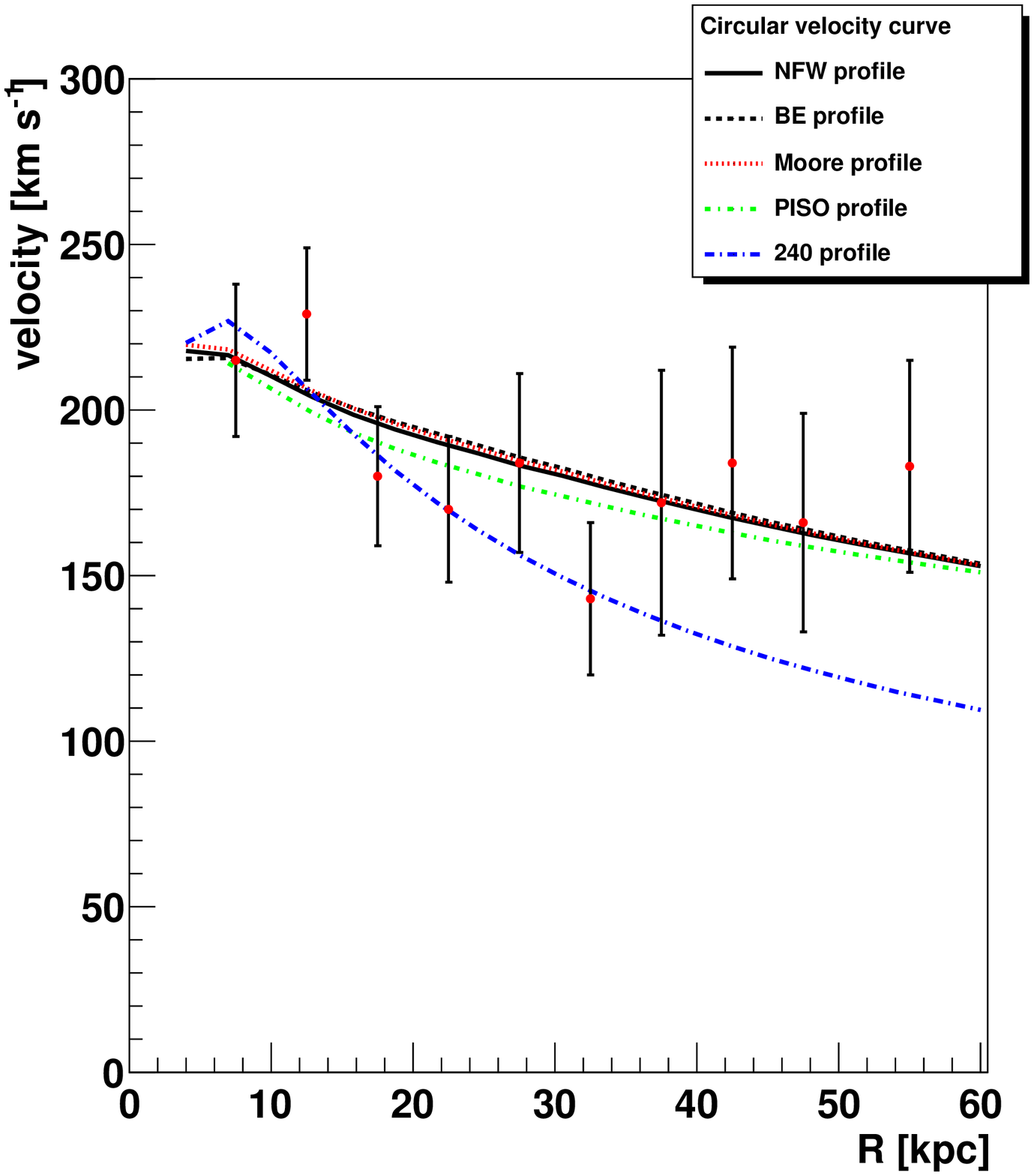}}
    \caption{The circular velocity curve for halo stars above a height of z $>$ 4 kpc is
             shown. On the left side the circular velocity curve for the NFW
	     profile calculated for different angles with respect to the Galactic
	     disc is shown. On the right
	     side the averaged circular velocity curves for the different halo profiles
	     are shown. The experimental data were obtained from the publication
         by \citet{Xue:2008se}.}
    \label{fig:rc_4kpc}
  \end{center}
\end{figure*}
\subsection{Rotation curve}
\label{subsec:RC}
Each object, which is bound to the MW, is orbiting around the
GC. Most of the
stars and the interstellar medium (gas, dust, etc.) are rotating
with a velocity distribution v(r). This velocity distribution is called the
{\it rotation curve} (RC) of the MW.\\
For a circular rotation of the objects
within the Galactic disc the rotation velocity is given by
the equality of the centripetal and the gravitational force
\begin{equation}
\frac{m v^2}{r} = m \cdot \frac{d \Phi}{dr},
\label{eq:CentripedalForce}
\end{equation}
where v is the rotation velocity at the Galactocentric distance r. The
gravitational potential $\Phi$ is given by the Poisson equation
\begin{equation}
\Delta \Phi = - 4 \pi G \rho (r),
\label{eq:PoissonEquation}
\end{equation}
where G is the gravitational constant and $\rho (\mathrm{r})$ is the
matter density distribution model. For a given DM density
profile the gravi\-tational potential and therefore the rotation
curve can be calculated and a comparison  with experimental data
is an important check for the
DM density profile.

The rotation curve can be most easily measured from the Doppler
shifts, like the 21 cm from neutral hydrogen or the rotational
transition lines of carbon monoxide (CO) in the millimeter wave
range. A review of the various methods was given by
\citet{Sofue:2000jx}. A recent summary of all data on the
rotation curve of the Milky Way can be found in the publication by
\citet{Sofue:2008wt}, where the numerical values can be found
on the author's webpage. For radii inside the solar radius the
distances do not need to be determined, since the maximum
Doppler shift observed in a given direction is in the
tangential direction of the circle, so the longitude of that
direction determines the distance from the center. For radii
larger than the solar radius the distances to the tracers have
to be determined independently, usually by the angular
thickness of the HI layer, as first proposed by
\citet{Merrifield:1992nb}. These independent distance
determinations lead to larger errors in the outer rotation
curve.

The most precise determination and with it the normalization of
the rotation curve is obtained from the Oort constants, which can
be determined from the precise distances and velocities of
nearby stars, as discussed in most textbooks, e.g. \citep{Binney,Sparke,Zeilik}.
These constants are defined as:
\begin{eqnarray}
\label{oort}
A & \equiv & -\frac{1}{2}\left[\frac{dv}{dr}|r_\odot -\frac{v_\odot}{r_\odot}\right]\approx 14.4\pm1.2\ \mathrm{km}\ \mathrm{s}^{-1}\ \mathrm{kpc}^{-1}\\
B & \equiv & -\frac{1}{2}\left[\frac{dv}{dr}|r_\odot +\frac{v_\odot}{r_\odot}\right]\approx -12.0\pm2.8\ \mathrm{km}\ \mathrm{s}^{-1}\ \mathrm{kpc}^{-1},
\end{eqnarray}
where the experimental values have been taken from
\citet{Kerr:1986hz}. One observes that
A-B $= {\mathrm{v}_\odot}/{\mathrm{r}_\odot}$ defines the local normalization of
the rotation curve, while A+B defines the slope at the solar
position.

The combination A-B can be more precisely determined than the individual constants.
\citet{Kerr:1986hz} found
$\mathrm{A-B} = \mathrm{v}_\odot/\mathrm{r}_\odot =$ 27$\pm$ 2.5  km s$^{-1}$ kpc$^{-1}$.
Using the proper motion of the black hole in the Galactic centre (Sgr A*)
 \citet{Reid:2004rd} found
\begin{equation}
A-B=v_\odot/r_\odot=29.45\pm 0.15\ \mathrm{km}\ \mathrm{s}^{-1}\ \mathrm{kpc}^{-1},
\label{oort1}
\end{equation}
in excellent agreement with recent measurements of parallaxes using the Very Large Baseline Interferometry
(VLBI) \citep{Reid:2009nj}, which yield A-B $= $v$_\odot$/r$_\odot =$ 30.3 $\pm$ 0.9  km s$^{-1}$ kpc$^{-1}$.
A further interesting property of the RC is its slope at the position of the Sun.
This value is strongly connected to the values of A and B and was determined
in the publication by \citet{Fuchs:2009mk} to be
\begin{equation}
  RC_{Slope} = \frac{\partial ln(v_\odot)}{\partial ln(r)} = -\frac{A+B}{A-B} = -0.006 \pm 0.016.
\end{equation}

From the velocities of stars orbiting Sgr A*, which is considered to be the centre of the Galaxy
 because of its small own velocity, the distance between the Sun and the GC has been determined to 
\citep{Gillessen:2008qv}:
 \begin{equation}
r_\odot=8.33\pm0.35\ \mathrm{kpc},
\label{r0}
\end{equation}
in agreement with previous authors \citep{Ghez:2008ms}.
With this Galactocentric distance one finds from Eq. 
\ref{oort1} a rotation velocity of the Sun
\begin{equation}
244 \pm 10\ \mathrm{km}\ \mathrm{s}^{-1},
\label{vsun}
\end{equation}
which is consistent with recent observations of Galactic masers
in \citet{Bovy:2009dr}, who used 
data from the Very Long Baseline Array (VLBA) and the 
Japanese VLBI Exploration of Radio Astronomy (VERA). 
This speed determines the mass of the Galaxy inside the solar radius.

In this analysis two different rotation velocities are
considered: the RC within the Galactic disc (Fig. \ref{fig:rc_simple})
and the velocity distribution for stars outside the disc with
z $>$ 4 kpc (Fig. \ref{fig:rc_4kpc}). They are discussed separately.
\subsubsection{Rotation curve in the disc}
For the RC within the disc a combination
of different measurements with different tracers has been summarized
 by \citet{Sofue:2008wt}. The
experimental data, which can be found on the author's web page,
were scaled to v$_\odot =$ 244 km\ s$^{-1}$ at a Galactocentric distance
of 8.3 kpc. Furthermore,  the rotation velocity was averaged
in 17 radial bins from the GC to a radius of 22 kpc, as shown
in Fig. \ref{fig:rc_simple}b and tabulated in Table
\ref{tab:velocities}. The shape of the measured
velocity distribution shows a strong increase of the rotation
velocity in the inner part of the Galaxy which presumably
results from the dense core of the Galaxy. For the inner Galaxy
the rotation curve is dominated by the visible matter and the
parametrization of section \ref{subsec:para_lumi} yields a
reasonable
description. However, at the outer Galaxy the experimental data
cannot be explained: all profiles predict a slow decrease of
the rotation velocity in contrast to the data, which show first
a decrease bet\-ween 6 and 10 kpc and then increases again.
Such a peculiar change of slope cannot be explained by a
smoothly decreasing DM density profile, but needs substructure,
e.g. the infall of a dwarf Galaxy, as mentioned in the
introduction and \citet{deBoer:2005tm}. Such a ringlike
substructure is supported by  the gas flaring
\citep{Kalberla:2007sr}. The thickness of the substructure is
of the order of 1 kpc, so it should not show up for halo stars
well above this height; this is indeed the case, as shown in
Fig. \ref{fig:rc_4kpc}, which will be discussed in the next
section.

\begin{table}[tbp]
  \begin{center}
    \caption{Averaged values of the Galactocentric distance
	     and the rotation velocity shown in Fig. \ref{fig:rc_simple}.}
    \begin{tabular}{l||c|c|c|c|c|c}\hline
Nr & r$_{\mathrm{Min}}$ & r$_{\mathrm{Max}}$ & r [kpc] & $\sigma_{\mathrm{r}}\ [$kpc$]$ & v [km s$^{-1}$] & $\sigma_{\mathrm{v}}$ [km s$^{-1}$] \\\hline
1 & 0.0 & 1.0 & 0.146 & 0.002 & 281.514 & 0.851 \\
2 & 1.0 & 1.5 & 1.190 & 0.011 & 261.127 & 0.783 \\
3 & 1.5 & 2.0 & 1.670 & 0.016 & 247.721 & 0.838 \\
4 & 2.0 & 2.5 & 2.253 & 0.016 & 236.071 & 0.640 \\
5 & 2.5 & 3.0 & 2.770 & 0.020 & 233.645 & 0.474 \\
6 & 3.0 & 4.0 & 3.481 & 0.017 & 242.435 & 0.307 \\
7 & 4.0 & 5.0 & 4.457 & 0.021 & 257.475 & 0.307 \\
8 & 5.0 & 6.0 & 5.481 & 0.026 & 257.255 & 0.321 \\
9 & 6.0 & 7.0 & 6.464 & 0.027 & 261.968 & 0.318 \\
10 & 7.0 & 8.0 & 7.852 & 0.001 & 251.095 & 0.198 \\
11 & 8.0 & 8.5 & 8.202 & 0.009 & 243.078 & 0.352 \\
12 & 8.5 & 9.0 & 8.761 & 0.046 & 233.942 & 1.278 \\
13 & 9.0 & 10.0& 9.478 & 0.052 & 231.191 & 1.434 \\
14 & 10.0& 11.0& 10.578 & 0.199 & 238.947 & 4.963 \\
15 & 11.0& 12.0& 11.500 & 0.227 & 253.366 & 5.631 \\
16 & 12.0& 16.0& 12.912 & 0.115 & 253.384 & 2.837 \\
17 & 16.0& 22.0& 17.141 & 0.796 & 274.735 & 13.594 \\\hline
    \end{tabular}
    \label{tab:velocities}
  \end{center}
\end{table}

One may argue
about the large uncertainties in the outer rotation curve,
where the distance {\it and} the velocity have to be determined
in contrast to the inner rotation curve, where the tangent
method yields the distance from the maximum velocity
\citep{Binney}. The rotation curve can be flattened, e.g. by
decreasing the distance between the Sun and the GC, but then
r$_\odot \approx $ 7 kpc is needed \citep{Honma:1997cg}, which
is clearly outside the present errors given in Eq. \ref{r0}.
Also the peculiar change of slope near 10 kpc does not
disappear.  
It should be noted that  such a change of slope happens in other spiral 
galaxies as well, as can be seen from the compilation of rotation curves 
in Fig. 4 in \citet{Sofue:2000jx}.
For the smooth DM density profiles discussed in
this paper this feature will be neglected.

\subsubsection{Rotation curve in the halo}
 The data in Fig. \ref{fig:rc_4kpc} were obtained from  a large
sample of roughly 2400 blue horizontal-branch (BHB) tracer stars,
as detected in Sloan Digital Sky Survey (SDSS), with
Galactocentric distances up to about 60 kpc and vertical
heights of z $>$ 4 kpc. In order to connect the observable
values - line-of-sight velocity and distance - to the circular
velocity v(r)$ = \sqrt{\mathrm{r}\ \partial \Phi / \partial \mathrm{r}}$ the halo
star distribution function from N-body simulations of the Galaxy
with an NFW profile was used.

In order to compare our  DM density profiles with these data,
the velocity curve is
calculated for different angles with respect to the Galactic disc.
Then the results are averaged. In Fig. \ref{fig:rc_4kpc}
the averaged circular velocity curve and the velocity curves
for an inclination angle with the normal to the disc of 10$^{\circ}$,
45$^{\circ}$ and 80$^{\circ}$ are shown for the NFW profile. The averaged
circular velocity curves for the five other
spherical halo profiles discussed before are  shown in
Fig. \ref{fig:rc_4kpc}b.
The circular velocity distribution is consistent
with the cuspy halo profiles and the PISO profile. The 240 profile cannot
describe the velocity distribution at large radii because of the too steep
decrease of the density at large radii $(\propto$ 1/r$^4)$.

\subsection{Surface density and Oort limit}
\label{ss_surface}
\begin{table}[tbp]
  \begin{center}
    \caption{Contributions to the local surface density of baryonic matter.
             The total values in the last row include $\pm 1\sigma$ errors.}
    \begin{tabular}{c|c|c}\hline
      Contribution     & $\Sigma$ [M$_\odot$ pc$^{-2}$] & Reference \\\hline\hline
      Visible stars    & 35 $\pm$ 5 & \citet{Gilmore:1989mw}\\
		       & 27         & \citet{Gould:1995tf}\\ 	
		       & 30         & \citet{Zheng:2001wc}\\\hline
      Stellar remnants & 3 $\pm$ 1  & \citet{Mera:1998qg}\\\hline
      Interstellar gas & 8 $\pm$ 5  & \citet{Dame:1993mw}\\
                       & 13 - 14    & \citet{Olling:2001yt}\\\hline
      Total            & 35-58      &\\ \hline
    \end{tabular}
    \label{tab:SD_visible}
  \end{center}
\end{table}
\begin{figure}[tbp]
  \begin{center}
    \includegraphics[width=0.49\textwidth,angle=0]{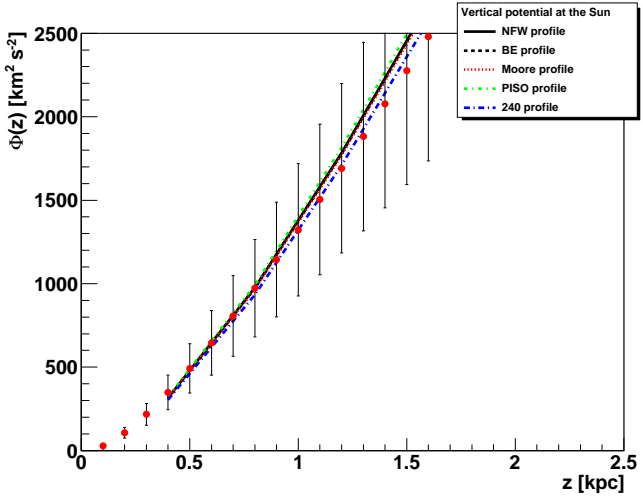}
    \caption{The vertical gravitational potential
             at the position of the Sun $\Phi(\mathrm{z})$.
              The experimental data is adapted from
             \citep{Bienayme:2005py} and the errors are assumed to be 30\%.}
    \label{fig:grav_pot_simple}
  \end{center}
\end{figure}
Jan Oort proposed and performed another interesting measurement:
from the star count  as function of their height above the disc
one obtains the local gravitational potential, which is directly
proportional to the mass in the plane of the MW.
Using the precise measurements from the Hipparcos satellite
\citet{Holmberg:2004fj} find for the local mass density, which
includes visible and dark matter,

\begin{equation}
\rho_{\odot,\mathrm{tot}} (z=0) = 0.102 \pm 0.010\ \mathrm{M}_\odot\ \mathrm{pc}^{-3}.
\label{oort2}
\end{equation}

This value was  determined by the precise star counts and
velocity measurements in a volume of 125 pc around the Sun by
the Hipparcos satellite. \citet{Korchagin:2003yk} analysed the
vertical potential at slightly larger distances (a vertical
cylinder of 200 pc radius and an extension of 400 pc out
of the Galactic plane). For the dynamical estimate of the
local volume density they obtain the same value with a
smaller error: $\rho_{\odot,\mathrm{tot}} (\mathrm{z}=0) =$ 0.100 $\pm$ 0.005
M$_{\odot}$ pc$^3$. To be conservative, we will use the value
from Eq. \ref{oort2}. It is called the Oort limit, since
it gives the lowest value for the local density.

Integrating the  density along the vertical direction
within \mbox{$\pm$ z} from the Galactic plane yields the surface density:
\begin{equation}
  \Sigma_{def}\ (< |z|) = \int\limits_{-z}^z \rho(z^\prime)\ dz^\prime.
\label{eq:SD_definition}
\end{equation}
\begin{table*}[tbp]
  \begin{center}
    \caption{Free and fixed parameters for the mass model of the
             Galaxy and experimental constraints. One observes that
             there are 5 free parameters and 8 constraints. Mass densities
             are in GeV cm$^{-3}$ or in M$_\odot$ pc$^{-3}$, where
             1 M$_\odot$ cm$^{-3} \equiv$ 37.97 GeV cm$^{-3}$.}
    \begin{tabular}{l|l||c|c|c}\hline
    \multicolumn{5}{c}{Free Parameters} \\\hline\hline
      Component & Parameter& Symbol & Value & Unit\\ \hline
      Halo&Local DM density& $\rho_{\odot,\mathrm{DM}}$& - & GeV/cm$^{-3}$ \\ \hline
      Halo&Scale Parameter& a& - & kpc \\ \hline
      Halo&eccentricity& $\epsilon_\mathrm{z}$& - & kpc \\ \hline
      Disc& Density at GC& $\rho_\mathrm{d}$ & - & GeV cm$^{-3}$ \\ \hline
      Disc& scale length&$r_\mathrm{d}$& - & kpc \\ \hline\hline
      \multicolumn{5}{c}{Fixed Parameters} \\\hline\hline
      Disc& scale height& $z_d$& 320 & pc \\ \hline
      Bulge& Density at GC& $\rho_\mathrm{b}$& 360& GeV cm$^{-3}$ \\ \hline
      Bulge& Eccentricity & $q_\mathrm{b}$ & 0.61 & \\ \hline
      Bulge& Scale   &      $r_\mathrm{t}$ & 0.6 & kpc \\ \hline
      Bulge& Scale   &      $r_{\mathrm{0,b}}$ & 0.7 & kpc \\ \hline
      Bulge& Slope   & $\gamma_\mathrm{b}$ & 1.6  & \\ \hline
      Bulge& Slope   & $ \beta_\mathrm{b}$ & 1.6  & \\ \hline
      \multicolumn{5}{c}{Constraints} \\\hline\hline
      All & Mass inside 60 kpc & M$_{\mathrm{R} < 60 \mathrm{kpc}}$ & 4.0 $\pm$ 0.7 & 10$^{\mathrm{11}}$ M$_\odot$ \\ \hline
      Local & Rotation speed Sun & v$_\odot$ & 244$\pm$ 10 & km s$^{-1}$ \\ \hline
      Local & Distance Sun-GC & r$_\odot$ &  8.33 $\pm$ 0.35 & kpc \\ \hline
      Local & Total Surface Density & $\Sigma_{|\mathrm{z}|<\mathrm{1.1}}$ & 71 $\pm$ 6 & M$_\odot$ pc$^{-2}$ \\ \hline
      Local & Visible Surface Density & $\Sigma_{vis}$& 48 $\pm$ 9 & M$_\odot$ pc$^{-2}$ \\ \hline
      Local & Mass Density &$\rho_{tot}$& 0.102 $\pm$ 0.01 & M$_\odot$ pc$^{-3}$ \\ \hline
      Local & Oort Constants & A-B & 29.45 $\pm$ 0.15 & km s$^{-2}$ kpc$^{-1}$ \\\hline
      Local & Slope of rotation curve & $\partial$ ln(v$_\odot$)/$\partial$ ln(r) & -0.006 $\pm$ 0.016 & \\\hline
    \end{tabular}
    \label{param}
  \end{center}
\end{table*}

\begin{table*}
\begin{center}
\caption{Fit results. The units of the different values are given in Table \ref{param}. 
         The $\chi^2$ contributions are given below the variable value in brackets.}
\label{tab:results}
{\scriptsize
\begin{tabular}{l|l|c|c|c|c||c|c|c|c|c|c|c|c|c|c}
\hline
 & &\multicolumn{4}{|c||}{Fitted Parameters}&\multicolumn{9}{c|}{Derived Quantities} & \\\hline\hline
Nr & Profile & $\rho_{\odot,\mathrm{DM}}$ & a & $\rho_\mathrm{d}$ & r$_\mathrm{d}$ & c$_{\mathrm{vir}}$ & $\rho_{\odot,\mathrm{tot}}$ & v$_\odot$ & M$_{\mathrm{60}}$ & M$_{\mathrm{tot}}$ & A-B & $\frac{\partial ln(v_\odot)}{\partial ln(r)}$ & $\Sigma_{|\mathrm{z}|<\mathrm{1.1}}$ & $\Sigma_{\mathrm{vis}}$ & $\chi^2$  \\\hline
1 & {NFW} & 0.32 $\pm$ 0.05 & 10 & 88.3 $\pm$ 19.8 & 2.5 $\pm$ 0.2 & 17.5 & 0.094 & 244.5 & 3.9 $\cdot$ 10$^{\mathrm{11}}$ & 6.5 $\cdot$ 10$^{\mathrm{11}}$ & 29.45 & -0.002 & 72.2 & 53.8 & \\
      & & & & & (0.13) & & (0.58) & (0.0) & (0.0) & & (0.01) & (0.06) & (0.04) & (0.41) & (1.25) \\\hline
2& {NFW} & 0.27 $\pm$ 0.06 & 15 & 113.1 $\pm$ 17.1 & 2.4 $\pm$ 0.1 & 12.0 & 0.096 & 244.5 & 4.0 $\cdot$ 10$^{\mathrm{11}}$ & 7.4 $\cdot$ 10$^{\mathrm{11}}$ & 29.45 & -0.003 & 71.1 & 55.4 & \\
      & & & & & (0.01) & & (0.38) & (0.0) & (0.0) & & (0.0) & (0.04) & (0.0) & (0.68) & (1.13)\\\hline
3 & {NFW} & 0.23 $\pm$ 0.05 & 20 & 128.7 $\pm$ 69.8 & 2.3 $\pm$ 0.1 & 9.5 & 0.097 & 244.4 & 4.1 $\cdot$ 10$^{\mathrm{11}}$ & 8.3 $\cdot$ 10$^{\mathrm{11}}$ & 29.45 & -0.003 & 70.2 & 56.6 & \\
      & & & & & (0.0) & & (0.27) & (0.0) & (0.02) & & (0.0) & (0.03) & (0.02) & (0.92) & (1.27)\\\hline\hline
4 & {NFW} & 0.32 $\pm$ 0.04 & 10.8 $\pm$ 3.4 & 91.0 $\pm$ 8.0 & 2.5 & 16.2 & 0.095 & 244.4 & 4.0 $\cdot$ 10$^{\mathrm{11}}$ & 6.7 $\cdot$ 10$^{\mathrm{11}}$ & 29.45 & -0.003 & 72.4 & 54.2 & \\
      & & & & & & & (0.50) & (0.0) & (0.0) & & (0.0) & (0.04) & (0.06) & (0.48) & (1.07) \\\hline
5 & {NFW} & 0.35 $\pm$ 0.06 & 14.9 $\pm$ 4.8 & 89.5 $\pm$ 8.2 & 2.5 & 13.4 & 0.094 & 244.4 & 4.25 $\cdot$ 10$^{\mathrm{11}}$ & 9.8 $\cdot$ 10$^{\mathrm{11}}$ & 29.45 & -0.003 & 73.7 & 53.2 & \\
      & & & & & & & (0.57) & (0.0) & & (0.0) & (0.0) & (0.05) & (0.21) & (0.34) & (1.17) \\\hline
6 & {NFW} & 0.39 $\pm$ 0.05 & 20.4 $\pm$ 4.5 & 88.1 $\pm$ 8.7 & 2.5 & 11.5 & 0.094 & 244.4 & 5.5 $\cdot$ 10$^{\mathrm{11}}$ & 1.49 $\cdot$ 10$^{\mathrm{12}}$ & 29.44 & -0.001 & 75.0 & 52.4 & \\
      & & & & & & & (0.62) & (0.0) & & (0.0) & (0.08) & (0.0) & (0.45) & (0.24) & (1.40) \\\hline
7 & {NFW} & 0.41 $\pm$ 0.05 & 25.2 $\pm$ 4.6 & 87.3 $\pm$ 8.9 & 2.5 & 10.1 & 0.094 & 244.4 & 6.5 $\cdot$ 10$^{\mathrm{11}}$ & 1.98 $\cdot$ 10$^{\mathrm{12}}$ & 29.44 & -0.001 & 75.9 & 51.9 & \\
      & & & & & & & (0.64) & (0.0) & & (0.0) & (0.0) & (0.09) & (0.68) & (0.19) & (1.61) \\\hline\hline
8 & {BE} & 0.25 $\pm$ 0.05 & 10.2 & 133.6 $\pm$ 15.2 & 2.29 $\pm$ 0.09 & 17.6 & 0.096 & 244.4 & 4.1 $\cdot$ 10$^{\mathrm{11}}$ & 7.5 $\cdot$ 10$^{\mathrm{11}}$ & 29.45 & -0.003 & 70.6 & 56.2 & \\
      & & & & & (0.0) & & (0.31) & (0.0) & (0.01) & & (0.0) & (0.03) & (0.01) & (0.84) & (1.20) \\\hline
9 & {Moore} & 0.25 $\pm$ 0.05 & 30.0 & 114.7 $\pm$ 17.3 & 2.36 $\pm$ 0.11 & 6.2 & 0.095 & 244.4 & 4.1 $\cdot$ 10$^{\mathrm{11}}$ & 7.6 $\cdot$ 10$^{\mathrm{11}}$ & 29.45 & -0.003 & 70.4 & 56.2 & \\
      & & & & & (0.01) & & (0.32) & (0.0) & (0.01) & & (0.0) & (0.04) & (0.01) & (0.82) & (1.21) \\\hline
10 & {PISO} & 0.20 $\pm$ 0.04 & 5.0 & 150.4 $\pm$ 12.9 & 2.21 $\pm$ 0.08 & 46 & 0.098 & 244.4 & 4.1 $\cdot$ 10$^{\mathrm{11}}$ & 1.45 $\cdot$ 10$^{\mathrm{12}}$ & 29.45 & -0.004 & 69.3 & 57.8 & \\
      & & & & & (0.02) & & (0.19) & (0.0) & (0.03) & & (0.0) & (0.02) & (0.08) & (1.18) & (1.54)  \\\hline
11 & {240} & 0.26 $\pm$ 0.03 & 4.0 & 53.1 $\pm$ 9.6 & 3.0 $\pm$ 0.2 & 26.3 & 0.095 & 244.6 & 1.7 $\cdot$ 10$^{\mathrm{11}}$ & 1.8 $\cdot$ 10$^{\mathrm{11}}$ & 29.47 & -0.002 & 70.0 & 55.0 & \\
      & & & & & (1.36) & & (0.51) & (0.0) & (10.45) & & (0.02) & (0.06) & (0.03) & (0.61) & (13.04) \\\hline\hline
12 & {NFW} & 0.52 $\pm$ 0.07 & 26.6 $\pm$ 4.9 & 84.2 $\pm$ 9.5 & 2.5 & 9.6 & 0.094 & 244.3 & 6.6 $\cdot$ 10$^{\mathrm{11}}$ & 1.95 $\cdot$ 10$^{\mathrm{12}}$ & 29.43 & -0.001 & 80.3 & 50.1 & \\
      & & & & & & & (0.66) & (0.01) & & (0.03) & (0.02) & (0.10) & (2.38) & (0.66) & (3.24) \\\hline
\end{tabular}
}
\end{center}
\end{table*}
It can be calculated either by integrating the matter density distribution
directly or by using the gravitational potential of the Galaxy. The integration
limit is conventionally defined to be 1.1 kpc.

According to Eq.
(\ref{eq:PoissonEquation}) this definition is equal to
\begin{eqnarray}
  \nonumber \Sigma_{pot}\ (< |z|) & = & \frac{1}{2 \pi G} \cdot \int\limits_{0}^z
  \Delta \Phi (r,\varphi,z^\prime)\ dz^\prime\\
  \nonumber & = & \frac{1}{2 \pi G} \cdot \int\limits_{0}^z \left ( \frac{1}{r}
  \frac{\partial}{\partial r} \left( r \frac{
  \partial \Phi}{\partial r} \right) + \frac{1}{r^2}\frac{\partial^2
  \Phi}{\partial \varphi^2} + \frac{\partial^2 \Phi}{\partial z^{\prime 2}}
  \right )\ dz^{\prime}\\
  \nonumber & = & \frac{1}{2 \pi G} \cdot \left ( \frac{\partial \Phi}
  {\partial z} + \int\limits_{0}^z \left ( \frac{1}{r}
  \frac{\partial}{\partial r} \left( r \frac{
  \partial \Phi}{\partial r} \right) + \frac{1}{r^2}\frac{\partial^2
  \Phi}{\partial \varphi^2} \right )\ dz^\prime \right )\\
  \nonumber & = & \frac{1}{2 \pi G} \cdot \left (
  \frac{\partial \Phi}{\partial z} + 2 (A^2 - B^2) |z| \right )\\
  & \approx & \frac{1}{2 \pi G} \cdot \left( \frac{\partial \Phi}{\partial z} \right),
\label{eq:SD_phi}
\end{eqnarray}
where A and B are the  Oort constants discussed before. Since
they are of the same order of magnitude, the surface density
integrated to z is proportional to the derivative of the potential
at height z, as shown by the last approximation in Eq.  \ref{eq:SD_phi}.

First the surface density of the luminous matter is considered.
Its experimental value is determined by the summation of the
different contributions to the luminous matter -
the stellar population, stellar remnants and the interstellar gas.
A summary of the different measurements was given
 by \citet{Naab:2005km}.
 The surface density of
the baryonic matter lies between 35 and 58 M$_\odot$ pc$^{-2}$
(Table \ref{tab:SD_visible}), which agrees with 48 $\pm$ 9
M$_\odot$ pc$^{-2}$ as estimated in \citet{Kuijken:1991mw}
and \citet{Holmberg:2004fj}.

Unfortunately, our local neighbourhood is not representative for the disc, since
we live in a  local underdensity - the
local bubble - with an extension of a few hundred pc, which could have been caused
by a series of rather recent SN explosions \citep{MaizApellaniz:2001vm}.
Therefore, a fit for the parameters of a mass model of the Galaxy might
have a somewhat higher surface density than the locally observed value.

The total surface density at the position of the Sun was
determined by \citet{Kuijken:1991mw} to be
\begin{equation}
\Sigma (< 1.1\ \mathrm{kpc}) = 71 \pm 6\ \mathrm{M}_\odot \mathrm{pc}^{-2}
\label{eq:sigma}
\end{equation}
from a parametrization of
a mass distribution. In the paper
by \citet{Holmberg:2004fj} the modeling of the vertical gravitational
potential resulted in $\Sigma (< 1.1\ \mathrm{kpc}) =$ 74 $\pm$ 6 M$_\odot$ pc$^{-2}$.
The most recent determination of the surface density by \citet{Bienayme:2005py}
is consistent with both measurements but allows somewhat larger errors:
$\Sigma (< 1.1\ \mathrm{kpc}) =$ 68 $\pm$ 11 M$_\odot$ pc$^{-2}$.
Comparing this total surface density with the baryonic surface density
shown in Table \ref{tab:SD_visible}, it is clear that the local
 gravitational potential is largely determined by the stars, so it will
be practically independent of the DM halo profile, as shown in Fig.
\ref{fig:grav_pot_simple} for different halo profiles.

%
\section{Numerical Determination of the Mass Model of the Galaxy}
\label{sec:fit}

\begin{figure*}[tbp]
  \begin{center}
    \includegraphics[width=0.7\textwidth,angle=0]{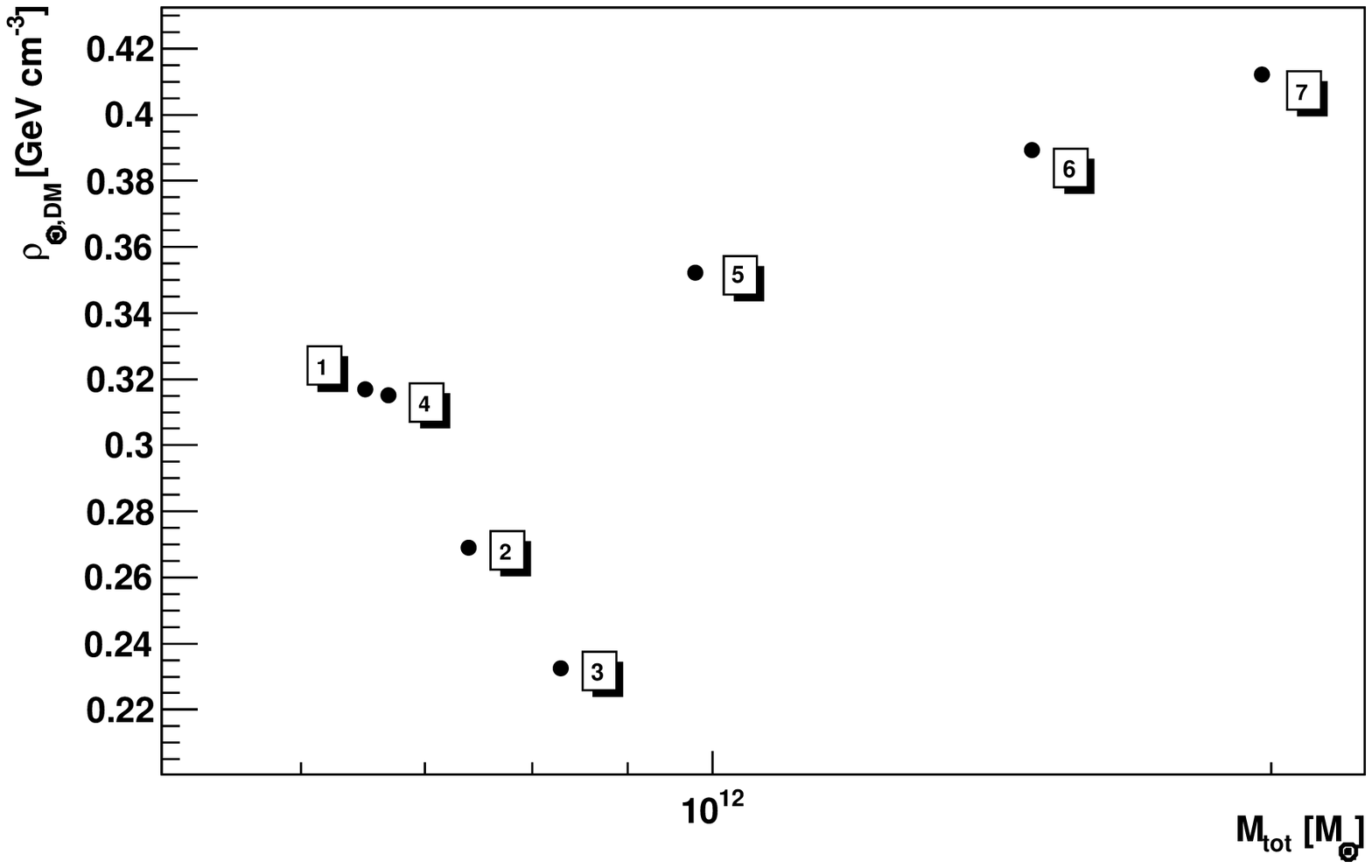}
    \caption{The local DM densities $\rho_{\odot,\mathrm{DM}}$ are shown for
	     different fits with different parameters.  The numbers correspond to
	     the numbers of the fit results in Table \ref{tab:results}.}
    \label{fig:diffuse_simple}
  \end{center}
\end{figure*}
As mentioned before, the three most important constraints for
the mass model of the Galaxy are given by the rotation curve
and the value v$_\odot$, the total mass M$_{\mathrm{tot}}$ and the local
mass density $\rho_{\odot,\mathrm{tot}}$. This can be easily seen as
follows: v$_\odot^2=$ v$_{\mathrm{vis}}^2 +$ v$_{\mathrm{DM}}^2$, 
where v$_{\mathrm{vis}}^2$ and
v$_{\mathrm{DM}}^2$ are proportional to $\rho_{\mathrm{vis}}$ and 
$\rho_{\mathrm{DM}}$,
respectively; for a given halo profile M$_{\mathrm{tot}}$ is determined
by $\rho_{\mathrm{DM}}$, while the Oort limit $\rho_{\odot,\mathrm{tot}}$ determines
$\rho_{\odot,\mathrm{tot}} = \rho_{\mathrm{vis}}+\rho_{\mathrm{DM}}$. So in principle one has
3 constraints with only 2 variables $\rho_{\mathrm{vis}}$ and
$\rho_{\mathrm{DM}}$, {\it if} the shapes  of the DM halo and the
visible matter would be known.

Unfortunately, additional important parameters are i) the
eccentricity of  the DM halo ii) the concentration of the DM
halo iii) the scale length of the disc iv) the mass in the
bar/bulge. In addition the mass model is sensitive to the
geometry, i.e. the Galactocentric distance from the Sun
d$_\odot$ and the halo profile. Additional constraints come
from the surface density, but here the visible surface density
has a large uncertainty as discussed before. The parameters and
constraints have been summarized in Table \ref{param}. The
parametrization of the mass of the bulge was chosen to describe
the rotation curve at small radii, which works reasonably well,
as can be
seen from Fig. \ref{fig:rc_simple}. Given that the mass model
is not very sensitive to this inner region, the parameters of
the bulge will not be varied anymore.

To optimize the remaining parameters in order to best describe the
data, the following $\chi^2$ function was minimized using the Minuit
package \citep{James:1975dr}
\begin{eqnarray}
\nonumber \chi^2 & = & \frac{(M_{tot}^{calc}-D)^2}{\sigma_{M_{tot}}^2}
      + \frac{(v_{\odot}^{calc}-D)^2}{\sigma_{v_{\odot}}^2}
      + \frac{(\rho_{tot}^{calc}-D)^2}{\sigma_{\rho_{tot}}^2} +\\
\nonumber & &     \frac{(\Sigma_{vis}^{calc}-D)^2}{\sigma_{\Sigma_{vis}}^2}
       + \frac{(\Sigma_{tot}^{calc}-D)^2}{\sigma_{\Sigma_{tot}}^2}
       + \frac{(r_{d}^{calc}-D)^2}{\sigma_{r_{d}}^2} +\\
& &      \frac{(RC_{Slope}^{calc}-D)^2}{\sigma_{RC_{Slope}}^2}
          + \frac{((A-B)^{calc}-D)^2}{\sigma_{A-B}^2}
\label{chi2}
\end{eqnarray}
The index $^{calc}$ means the observables were calculated from
the fitted parameters, while D  denotes the experimental data
for the observable and $\sigma$ its error. The constraints
have been summarized in Table \ref{param}.

The fit shows a more than 95\% positive correlation between the
local dark matter density and the scale length of DM halo a
and an equally large negative correlation with the scale length
r$_\mathrm{d}$ of the baryonic disc. Consequently, it is difficult to
leave parameters free in the fit. Therefore the fit was
first performed for fixed values of a (rows 1-3 of Table \ref{tab:results})
and then r$_\mathrm{d}$ was fixed (rows
4-7). With the other free parameters all experimental
constraints could be met, as indicated by the $\chi^2$ values
in brackets below the fitted values in Table \ref{tab:results}. Of
course, the total mass changed for the different fits. Fig.
\ref{fig:diffuse_simple} shows the resulting local DM density versus the total
mass, as calculated from the fitted parameters. It shows that
in spite of the small errors  for the local density in
individual fits  the spread in density is still quite large.

The fit was repeated for other halo profiles, which gave
simi\-larly good $\chi^2$ values, as shown by  rows 9-11 in Table
\ref{tab:results}. So with the present data one cannot distinguish
the different halo profiles.

Sofar only spherical halos have been discussed. Allowing oblate
halos  with a ratio of short-to-long axis of 0.7 the local DM
density increases by about 20\%, as shown by the last row of
Table \ref{tab:results}. As mentioned before, dark discs can
enhance this value considerably more, so the uncertainty
usually quoted for the local dark matter density in the range
of 0.2 to 0.7 GeV cm$^{-3}$ (0.005 - 0.018 M$_\odot$ pc$^{-3}$)
\citep{Amsler:2008zzb,Gates:1995dw} is still valid
in spite of the considerably improved data.


\section{Conclusion}
\label{sec:Conclusion}
In this analysis five different halo profiles are compared with recent
dynamical constraints as summarized in Table \ref{param}.
The change of slope in the RC around 10 kpc (Fig. \ref{fig:rc_simple})
was ignored, so the monotonical decreasing RC 
for the smooth halo profiles do not describe the data well. 
The change of slope may be related to a
ringlike DM substructure , as indicated by the structure in the
gas flaring \citep{Kalberla:2007sr} and by the structure in the
diffuse gamma radiation \citep{deBoer:2005tm}.
Such a ringlike structure of DM gives a perfect description of the rotation
curve, especially the fast decrease between 6 and 10 kpc. If the DM substructure
is included, the local DM density increases above the values found in this analysis,
so the  values quoted here should be considered lower limits.

The astronomical constraints are
consistent with a density model of the Galaxy consisting of a
central bulge, a disc and an extended DM halo with a cuspy
density profile and a local DM density between 0.2 GeV
cm$^{-3}$ (0.005 M$_\odot$ pc$^{-3}$) and 
0.4 GeV cm$^{-3}$ (0.01 M$_\odot$ pc$^{-3}$), as shown in Fig.
\ref{fig:diffuse_simple}. Strong positive and negative
correlations between the parameters were found in the fit
and they are causing the obvious correlations between
$\rho_{\odot,\mathrm{DM}}$ and M$_{\mathrm{tot}}$ in Fig. \ref{fig:diffuse_simple}.
For non-spherical haloes these values can be enhanced by
20\%. If dark discs are considered, densities up 
to 0.7 GeV cm$^{-3}$ (0.018 M$_\odot$ pc$^{-3}$)
can be easily imagined, so the previous quoted range of
0.2 - 0.7 GeV cm$^{-3}$ (0.005 - 0.018 M$_\odot$ pc$^{-3}$)
seems still valid.
This range is considerably larger than the values quoted by
analyses which used a Markov Chain method to minimize the
likelihood; they find $\rho_{\odot,\mathrm{DM}} = 0.39 \pm 
0.03$ GeV cm$^{-3}$ \citep{Catena:2009mf} and 
$\rho_{\odot,\mathrm{DM}} = 0.32 \pm 
0.07$ GeV cm$^{-3}$ \citep{Strigari:2009zb} respectively.
But given the good $\chi^2$ values for our fits obtained for a 
large range of DM densities we 
see no way that the errors can be as small as quoted by these authors.

\bibliographystyle{hapalike}
\bibliography{aa}

\begin{thebibliography}{}

\bibitem[Amsler et~al., 2008]{Amsler:2008zzb}
Amsler, C. et~al. (2008).
\newblock {Review of particle physics}.
\newblock {\em Phys. Lett.}, B667:1.

\bibitem[Battaglia et~al., 2005]{Battaglia:2005rj}
Battaglia, G. et~al. (2005).
\newblock {The radial velocity dispersion profile of the Galactic halo:
  Constraining the density profile of the dark halo of the Milky Way}.
\newblock {\em Mon. Not. Roy. Astron. Soc.}, 364:433--442, astro-ph/0506102.

\bibitem[Battaglia et~al., 2006]{Battaglia1:2005rj}
Battaglia, G. et~al. (2006).
\newblock {The radial velocity dispersion profile of the Galactic halo:
  Constraining the density profile of the dark halo of the Milky Way}.
\newblock {\em Mon. Not. Roy. Astron. Soc.}, 370:1055--1056, astro-ph/0506102.

\bibitem[Bekenstein and Milgrom, 1984]{Bekenstein:1984tv}
Bekenstein, J. and Milgrom, M. (1984).
\newblock {Does the missing mass problem signal the breakdown of Newtonian
  gravity?}
\newblock {\em Astrophys. J.}, 286:7--14.

\bibitem[Bienayme et~al., 2009]{Bienayme:2009wb}
Bienayme, O., Famaey, B., Wu, X., Zhao, H.~S., and Aubert, D. (2009).
\newblock {Galactic kinematics with modified Newtonian dynamics}.
\newblock 0904.3893.

\bibitem[Bienayme et~al., 2005]{Bienayme:2005py}
Bienayme, O., Soubiran, C., Mishenina, T.~V., Kovtyukh, V.~V., and Siebert, A.
  (2005).
\newblock {Vertical distribution of Galactic disk stars: III. The Galactic disk
  surface mass density from red clump giants}.
\newblock astro-ph/0510431.

\bibitem[Binney and Evans, 2001]{Binney:2001wu}
Binney, J.~J. and Evans, N.~W. (2001).
\newblock {Cuspy Dark-Matter Haloes and the Galaxy}.
\newblock {\em Mon. Not. Roy. Astron. Soc.}, 327:L27, astro-ph/0108505.

\bibitem[Binney, 1998]{Binney}
Binney, E. \&~Merrifield, M.~R. (1998).
\newblock {\em {Galactic astronomy}}.
\newblock {Princeton University Press}.

\bibitem[Blumenthal et~al., 1986]{Blumenthal:1985qy}
Blumenthal, G.~R., Faber, S.~M., Flores, R., and Primack, J.~R. (1986).
\newblock {Contraction of Dark Matter Galactic Halos Due to Baryonic Infall}.
\newblock {\em Astrophys. J.}, 301:27.

\bibitem[Bovy et~al., 2009]{Bovy:2009dr}
Bovy, J., Hogg, D.~W., and Rix, H.-W. (2009).
\newblock {Galactic masers and the Milky Way circular velocity}.
\newblock {\em Astrophys. J.}, 704:1704--1709, 0907.5423.

\bibitem[Cardone and Sereno, 2005]{Cardone:2005qq}
Cardone, V.~F. and Sereno, M. (2005).
\newblock {Modelling the Milky Way through adiabatic compression of cold dark
  matter halo}.
\newblock astro-ph/0501567.

\bibitem[Catena and Ullio, 2009]{Catena:2009mf}
Catena, R. and Ullio, P. (2009).
\newblock {A novel determination of the local dark matter density}.
\newblock {\em submitted to JCAP}, 0907.0018.

\bibitem[Dame, 1993]{Dame:1993mw}
Dame, T.~M. (1993).
\newblock {The Distribution of Neutral Gas in the Milky Way}.
\newblock In Holt, S.~S. and Verter, F., editors, {\em Back to the Galaxy, AIP
  Conf. 278}, pages 267--278.

\bibitem[de~Boer et~al., 2005]{deBoer:2005tm}
de~Boer, W., Sander, C., Zhukov, V., Gladyshev, A.~V., and Kazakov, D.~I.
  (2005).
\newblock {EGRET excess of diffuse galactic gamma rays as tracer of dark
  matter}.
\newblock {\em Astron. Astrophys.}, 444:51, astro-ph/0508617.

\bibitem[Freudenreich, 1998]{Freudenreich:1997bx}
Freudenreich, H.~T. (1998).
\newblock {COBE's Galactic Bar and Disk}.
\newblock {\em Astrophys. J.}, 492:495--510, astro-ph/9707340.

\bibitem[Fuchs et~al., 2009]{Fuchs:2009mk}
Fuchs, B. et~al. (2009).
\newblock {The kinematics of late type stars in the solar cylinder studied with
  SDSS data}.
\newblock {\em Astron. J.}, 137:4149--4159, 0902.2324.

\bibitem[Gates et~al., 1995]{Gates:1995dw}
Gates, E.~I., Gyuk, G., and Turner, M.~S. (1995).
\newblock {The Local halo density}.
\newblock {\em Astrophys. J.}, 449:L123--L126, astro-ph/9505039.

\bibitem[Gentile et~al., 2007]{Gentile:2007sb}
Gentile, G., Tonini, C., and Salucci, P. (2007).
\newblock {LambdaCDM Halo Density Profiles: where do actual halos converge to
  NFW ones?}
\newblock {\em Astron. Astrophys.}, 467:925--931, astro-ph/0701550.

\bibitem[Ghez et~al., 2008]{Ghez:2008ms}
Ghez, A.~M. et~al. (2008).
\newblock {Measuring Distance and Properties of the Milky Way's Central
  Supermassive Black Hole with Stellar Orbits}.
\newblock 0808.2870.

\bibitem[Gillessen et~al., 2009]{Gillessen:2008qv}
Gillessen, S. et~al. (2009).
\newblock {Monitoring stellar orbits around the Massive Black Hole in the
  Galactic Center}.
\newblock {\em Astrophys. J.}, 692:1075--1109, 0810.4674.

\bibitem[Gilmore and Reid, 1983]{Gilmore:1983bv}
Gilmore, G. and Reid, N. (1983).
\newblock {New light on faint stars. III - Galactic structure towards the South
  Pole and the Galactic thick disc}.
\newblock {\em Mon. Not. Roy. Astron. Soc.}, 202:1025--1047.

\bibitem[Gilmore et~al., 1989]{Gilmore:1989mw}
Gilmore, G., Wyse, R. F.~G., and Kuijken, K. (1989).
\newblock {Stellar dynamics and Galactic evolution}.
\newblock In {\em Evolutionary phenomena in galaxies}, pages 172--200.
  Cambridge University Press.

\bibitem[Gould et~al., 1996]{Gould:1995tf}
Gould, A., Bahcall, J.~N., and Flynn, C. (1996).
\newblock {Disk M dwarf luminosity function from HST star counts}.
\newblock {\em Astrophys. J.}, 465:759, astro-ph/9505087.

\bibitem[Gunn, 1977]{Gunn:1977mw}
Gunn, J.~E. (1977).
\newblock {Massive galactic halos. I - Formation and evolution}.
\newblock {\em Astrophys. J.}, 218:592.

\bibitem[Hammer et~al., 2007]{Hammer:2007ki}
Hammer, F., Puech, M., Chemin, L., Flores, H., and Lehnert, M. (2007).
\newblock {The Milky Way: An Exceptionally Quiet Galaxy: Implications for the
  formation of spiral galaxies}.
\newblock {\em Astrophys. J.}, 662:322--334, astro-ph/0702585.

\bibitem[Holmberg and Flynn, 2004]{Holmberg:2004fj}
Holmberg, J. and Flynn, C. (2004).
\newblock {The local surface density of disc matter mapped by Hipparcos}.
\newblock {\em Mon. Not. Roy. Astron. Soc.}, 352:440, astro-ph/0405155.

\bibitem[Honma and Sofue, 1997]{Honma:1997cg}
Honma, M. and Sofue, Y. (1997).
\newblock {Mass of the galaxy inferred from outer rotation curve}.
\newblock astro-ph/9611156.

\bibitem[James and Roos, 1975]{James:1975dr}
James, F. and Roos, M. (1975).
\newblock {Minuit: A System for Function Minimization and Analysis of the
  Parameter Errors and Correlations}.
\newblock {\em Comput. Phys. Commun.}, 10:343--367.

\bibitem[Kalberla et~al., 2007]{Kalberla:2007sr}
Kalberla, P. M.~W., Dedes, L., Kerp, J., and Haud, U. (2007).
\newblock {Dark matter in the Milky Way, II. the HI gas distribution as a
  tracer of the gravitational potential}.
\newblock 0704.3925.

\bibitem[Kerr and Lynden-Bell, 1986]{Kerr:1986hz}
Kerr, F.~J. and Lynden-Bell, D. (1986).
\newblock {Review of galactic constants}.
\newblock {\em Mon. Not. Roy. Astron. Soc.}, 221:1023.

\bibitem[Klypin et~al., 2002]{Klypin:2001xu}
Klypin, A., Zhao, H., and Somerville, R.~S. (2002).
\newblock {LCDM-based models for the Milky Way and M31 I: Dynamical Models}.
\newblock {\em Astrophys. J.}, 573:597--613, astro-ph/0110390.

\bibitem[Korchagin et~al., 2003]{Korchagin:2003yk}
Korchagin, V.~I., Girard, T.~M., Borkova, T.~V., Dinescu, D.~I., and van
  Altena, W.~F. (2003).
\newblock {Local Surface Density of the Galactic Disk from a 3-D Stellar
  Velocity Sample}.
\newblock astro-ph/0308276.

\bibitem[Kroupa et~al., 1993]{Kroupa:1993ga}
Kroupa, P., Tout, C.~A., and Gilmore, G. (1993).
\newblock {The Distribution of low mass stars in the galactic disc}.
\newblock {\em Mon. Not. Roy. Astron. Soc.}, 262:545.

\bibitem[Kuijken and Gilmore, 1991]{Kuijken:1991mw}
Kuijken, K. and Gilmore, G. (1991).
\newblock {The galactic disk surface mass density and the Galactic force K(z)
  at Z = 1.1 kiloparsecs}.
\newblock {\em Astrophys. J.}, 367:L9--L13.

\bibitem[Ludlow et~al., 2009]{Ludlow:2008qf}
Ludlow, A.~D. et~al. (2009).
\newblock {The Unorthodox Orbits of Substructure Halos}.
\newblock {\em Astrophys. J.}, 692:931--941, 0801.1127.

\bibitem[Maiz-Apellaniz, 2001]{MaizApellaniz:2001vm}
Maiz-Apellaniz, J. (2001).
\newblock {The origin of the Local Bubble}.
\newblock astro-ph/0108472.

\bibitem[Mashchenko et~al., 2006]{Mashchenko:2006dm}
Mashchenko, S., Couchman, H. M.~P., and Wadsley, J. (2006).
\newblock {Cosmological puzzle resolved by stellar feedback in high redshift
  galaxies}.
\newblock {\em Nature}, 442:539, astro-ph/0605672.

\bibitem[Mera et~al., 1998]{Mera:1998qg}
Mera, D., Chabrier, G., and Schaeffer, R. (1998).
\newblock {Towards a consistent model of the Galaxy: I. kinematic properties,
  star counts and microlensing observations}.
\newblock astro-ph/9801051.

\bibitem[Merrifield, 1992]{Merrifield:1992nb}
Merrifield, M.~R. (1992).
\newblock {The Rotation curve of the milky way to 2.5-R(0) from the thickness
  of the HI layer}.
\newblock {\em Astron. J.}, 103.
\newblock CITA-91-44.

\bibitem[Moore et~al., 1999]{Moore:1999nt}
Moore, B. et~al. (1999).
\newblock {Dark matter substructure within galactic halos}.
\newblock {\em Astrophys. J.}, 524:L19--L22.

\bibitem[Naab and Ostriker, 2006]{Naab:2005km}
Naab, T. and Ostriker, J.~P. (2006).
\newblock {A simple model for the evolution of disc galaxies: The Milky Way}.
\newblock {\em Mon. Not. Roy. Astron. Soc.}, 366:899--917, astro-ph/0505594.

\bibitem[Navarro et~al., 1997]{Navarro:1996gj}
Navarro, J.~F., Frenk, C.~S., and White, S. D.~M. (1997).
\newblock {A Universal Density Profile from Hierarchical Clustering}.
\newblock {\em Astrophys. J.}, 490:493--508, astro-ph/9611107.

\bibitem[Oh et~al., 2008]{Oh:2008ww}
Oh, S.-H., de~Blok, W. J.~G., Walter, F., Brinks, E., and Kennicutt, Robert~C.,
  J. (2008).
\newblock {High-resolution dark matter density profiles of THINGS dwarf
  galaxies: Correcting for non-circular motions}.
\newblock 0810.2119.

\bibitem[Ojha et~al., 1996]{Ojha:1995ie}
Ojha, D., Bienayme, O., Robin, A., Creze, M., and Mohan, V. (1996).
\newblock {Structure and kinematical properties of the Galaxy at intermediate
  galactic latitudes}.
\newblock {\em Astron. Astrophys.}, 311:456--469, astro-ph/9511049.

\bibitem[Olling and Merrifield, 2001]{Olling:2001yt}
Olling, R.~P. and Merrifield, M.~R. (2001).
\newblock {Luminous and Dark Matter in the Milky Way}.
\newblock {\em Mon. Not. Roy. Astron. Soc.}, 326:164, astro-ph/0104465.

\bibitem[Purcell et~al., 2009]{Purcell:2009yp}
Purcell, C.~W., Bullock, J.~S., and Kaplinghat, M. (2009).
\newblock {The Dark Disk of the Milky Way}.
\newblock {\em Astrophys. J.}, 703:2275--2284, 0906.5348.

\bibitem[Reid, 2009]{Reid:2008fp}
Reid, M.~J. (2009).
\newblock {Is there a Supermassive Black Hole at the Center of the Milky Way?}
\newblock {\em Int. J. Mod. Phys.}, D18:889--910, 0808.2624.

\bibitem[Reid and Brunthaler, 2004]{Reid:2004rd}
Reid, M.~J. and Brunthaler, A. (2004).
\newblock {The Proper Motion of Sgr A*: II. The Mass of Sgr A*}.
\newblock {\em Astrophys. J.}, 616:872--884, astro-ph/0408107.

\bibitem[Reid et~al., 2009]{Reid:2009nj}
Reid, M.~J. et~al. (2009).
\newblock {Trigonometric Parallaxes of Massive Star Forming Regions: VI.
  Galactic Structure, Fundamental Parameters and Non- Circular Motions}.
\newblock {\em Astrophys. J.}, 700:137--148, 0902.3913.

\bibitem[Ricotti, 2003]{Ricotti:2002qu}
Ricotti, M. (2003).
\newblock {Dependence of the Inner DM Profile on the Halo Mass}.
\newblock {\em Mon. Not. Roy. Astron. Soc.}, 344:1237, astro-ph/0212146.

\bibitem[Robin et~al., 1996]{Robin:1995fr}
Robin, A.~C., Haywood, M., Creze, M., Ojha, D.~K., and Bienayme, O. (1996).
\newblock {The thick disc of the galaxy: Sequel of a merging event}.
\newblock {\em Astron. Astrophys.}, 305:125, astro-ph/9504090.

\bibitem[Sakamoto et~al., 2003]{Sakamoto:2002zr}
Sakamoto, T., Chiba, M., and Beers, T.~C. (2003).
\newblock {The Mass of the Milky Way: Limits from a Newly Assembled Set of Halo
  Objects}.
\newblock {\em Astron. Astrophys.}, 397:899--912, astro-ph/0210508.

\bibitem[Salucci et~al., 2007]{Salucci:2007tm}
Salucci, P. et~al. (2007).
\newblock {The universal rotation curve of spiral galaxies. II: The dark matter
  distribution out to the virial radius}.
\newblock {\em Mon. Not. Roy. Astron. Soc.}, 378:41--47, astro-ph/0703115.

\bibitem[Sofue et~al., 2008]{Sofue:2008wt}
Sofue, Y., Honma, M., and Omodaka, T. (2008).
\newblock {Unified Rotation Curve of the Galaxy -- Decomposition into de
  Vaucouleurs Bulge, Disk, Dark Halo, and the 9-kpc Rotation Dip --}.
\newblock 0811.0859.

\bibitem[Sofue and Rubin, 2001]{Sofue:2000jx}
Sofue, Y. and Rubin, V. (2001).
\newblock {Rotation Curves of Spiral Galaxies}.
\newblock {\em Ann. Rev. Astron. Astrophys.}, 39:137--174, astro-ph/0010594.

\bibitem[Sparke, 2007]{Sparke}
Sparke, L. S. \&~Gallagher, J.~S. (2007).
\newblock {\em {Galaxies in the Universe - An Introduction}}.
\newblock {Cambridge University Press}.

\bibitem[Spooner, 2007]{Spooner:2007zh}
Spooner, N.~J. (2007).
\newblock {Direct Dark Matter Searches}.
\newblock {\em J. Phys. Soc. Jap.}, 76:111016, 0705.3345.

\bibitem[Springel et~al., 2008a]{Springel:2008by}
Springel, V. et~al. (2008a).
\newblock {A blueprint for detecting supersymmetric dark matter in the Galactic
  halo}.
\newblock 0809.0894.

\bibitem[Springel et~al., 2008b]{Springel:2008zz}
Springel, V. et~al. (2008b).
\newblock {Prospects for detecting supersymmetric dark matter in the Galactic
  halo}.
\newblock {\em Nature}, 456N7218:73--80.

\bibitem[Strigari and Trotta, 2009]{Strigari:2009zb}
Strigari, L.~E. and Trotta, R. (2009).
\newblock {Reconstructing WIMP Properties in Direct Detection Experiments
  Including Galactic Dark Matter Distribution Uncertainties}.
\newblock {\em JCAP}, 0906.5361.
\newblock in press.

\bibitem[Wilkinson and Evans, 1999]{Wilkinson:1999hf}
Wilkinson, M.~I. and Evans, N.~W. (1999).
\newblock {The Present and Future Mass of the Milky Way Halo}.
\newblock {\em Mon. Not. Roy. Astron. Soc.}, 310:645, astro-ph/9906197.

\bibitem[Xue et~al., 2008]{Xue:2008se}
Xue, X.~X. et~al. (2008).
\newblock {The Milky Way's Circular Velocity Curve to 60 kpc and an Estimate of
  the Dark Matter Halo Mass from Kinematics of ~2400 SDSS Blue Horizontal
  Branch Stars}.
\newblock {\em Astrophys. J.}, 684:1143--1158, 0801.1232.

\bibitem[Zeilik, 1998]{Zeilik}
Zeilik, M. \&~Gregory, S. (1998).
\newblock {\em {Introductory Astronomy and Astrophysics}}.
\newblock {Saunders College}.

\bibitem[Zheng et~al., 2001]{Zheng:2001wc}
Zheng, Z., Flynn, C., Gould, A., Bahcall, J.~N., and Salim, S. (2001).
\newblock {M Dwarfs from Hubble Space Telescope Star Counts. IV}.
\newblock {\em Astrophys. J.}, 555:393--404, astro-ph/0102442.

\end{thebibliography}

\end{document}